# Investigation of broadband terahertz generation from metasurface

Ming Fang,[1,3] Kaikun Niu,[1] Zhiaxiang Huang, [1,*] Wei E. I. Sha, [2,5,*] Xianliang Wu, [1] Thomas Koschny, [3] and Costas M. Soukoulis[3,4]

[1]*Key Laboratory of Intelligent Computing and Signal Processing, Ministry of Education, Anhui University, Hefei 230039, China*
[2]*Key Laboratory of Micro-nano Electronic Devices and Smart Systems of Zhejiang Province, College of Information Science & Electronic Engineering, Zhejiang University, Hangzhou 310027, China*
[3]*Ames Laboratory, U.S. DOE and Department of Physics and Astronomy, Iowa State University Ames, IA 50011 , USA*
[4]*Institute of Electronic Structure and Lasers (IESL), FORTH, 71110 Heraklion , Crete , Greece*
[*]*weisha@zju.edu.cn* [*]*zhixianghuang@ahu.edu.cn*

**Abstract:** The nonlinear metamaterials have been shown to provide nonlinear properties with high nonlinear conversion efficiency and in a myriad of light manipulation. Here we study terahertz generation from nonlinear metasurface consisting of single layer nanoscale split-ring resonator array. The terahertz generation due to optical rectification by the second-order nonlinearity of the split-ring resonator is investigated by a time-domain implementation of the hydrodynamic model for electron dynamics in metal. The results show that the nonlinear metasurface enables us to generate broadband terahertz radiation and free from quasi-phase-matching conditions. The proposed scheme provides a new concept of broadband THz source and designing nonlinear plasmonic metamaterials.

## 1. Introduction

Metamaterials are artificial materials in which subwavelength units are the basic elements interacting with the electromagnetic field. They have been reported to be adept at providing a myriad of unconventional properties. These properties give potential applications of plasmonic metasurfaces, including nanoantennas, sensing, lasing and plasmonic devices [1-5]. The fields of metamaterials have made spectacular progress in recent years. Meanwhile, nonlinear phenomena in metamaterials have been actively reported. For instance, metamaterials composed of split-ring resonators (SRRs) have been found to have a strong second-harmonic generation (SHG) when it was excited resonantly, in comparison to other metallic nanostructures [6,7]. This is due to the resonant nonlinearity and strong local field enhancement from non-centrosymmetric SRRs. Furthermore, optical third-harmonic generation (THG) has been studied in fishnet metamaterials, resulting from natural harmonic emission of the electric dipole [8-10]. The nonlinear properties in plasmonic metamaterials arise from the intrinsic nonlinearities of metals which relate to the complex dynamics of free electrons. These works broaden the scope of nonlinear metamaterials.

Meanwhile, the interest in terahertz frequencies grew strongly due to the advances in terahertz generation and detection [11,12]. One of the challenge and research directions in terahertz technologies is to generate broadband terahertz radiation. In regard to the problem, using optical rectification in nonlinear crystals such as GaAs/GaP/GaSe/LiNbO3/ZnTe is the most common solution [13-17]. However, all of these nonlinear crystals have undesired gaps in the terahertz region and require quasi-phase-matching conditions.

In this paper, we investigate a single-layer metasurface THz emitter consisting of split-ring resonators (SRRs) with 50 nm thickness. Terahertz signals can be generated from the SRRs by exciting their magnetic-dipole modes at the infrared frequencies. To reveal the mechanism of THz generation from the metasurface, a time-domain implementation of the hydrodynamic model for electron dynamics in metals is used. Both linear and nonlinear dynamics of the electrons in metamaterials are fully considered. The rigorous method allows us to study the nonlinear processes from plasmonic metamaterials. The metasurface is a single layer and its thickness is thousand times thinner than optimal nonlinear crystals but free from the quasi-phase-matching conditions. Moreover, the generated terahertz emission spectrum is continuous and the metasurface THz emitter can easily be scaled to arbitrary operation frequencies by optimizing the dimensions of the SRR unit. Our results open unique opportunities for designing nonlinear metamaterials and terahertz emitters.

## 2. Theoretical Models

The interaction of electromagnetic fields **E** and **H** with metal can be described by Maxwell's equations and hydrodynamic model.

$$\nabla \times \mathbf{H} = \varepsilon_0 \frac{\partial \mathbf{E}}{\partial t} + \frac{\partial \mathbf{P}}{\partial t}, \tag{1}$$

$$\nabla \times \mathbf{E} = -\mu_0 \frac{\partial \mathbf{H}}{\partial t}, \tag{2}$$

$$\frac{\partial \mathbf{v}}{\partial t} + \mathbf{v} \cdot \nabla \mathbf{v} = -\frac{e}{m}(\mathbf{E} + \mu_0 \mathbf{v} \times \mathbf{H}) - \gamma \mathbf{v} - \frac{\nabla p}{n}, \tag{3}$$

$$\frac{\partial n}{\partial t} = -\nabla \cdot (n\mathbf{v}). \tag{4}$$

Here $\varepsilon_0$, $\mu_0$, $m$, $e$, and $\gamma$ are the vacuum permittivity, vacuum permeability, electron mass, electron charge and electron collision rate, respectively. $n(\mathbf{r},t)$ and $\mathbf{v}(\mathbf{r},t)$ are the time- and position-dependent electron density and velocity. $p$ in Eq. (3) is the electron pressure and evaluated by the Thomas-Fermi model $p = (3\pi^2)^{2/3}(\hbar/5)n^{5/3}$ [18]. The polarization P of materials in Eq. (1) can be related to the hydrodynamic variables by

$$\dot{\mathbf{P}} = \mathbf{J} = -en\mathbf{v}, \tag{5}$$

$$\rho = e(n_e - n_0). \tag{6}$$

Equations (1)-(5) provide a self-consistent description of the electron gas in metals. For numerical analyses of nonlinear responses in metals, the set of equations (1)-(5) of the multiphysics model can be implemented with the help of classic electromagnetic computational methods [19-23]. Here, we transform the microscopic hydrodynamic description into the classic electromagnetic frame. By substituting (5) and (6) into (3) and (4), we have

$$\frac{\partial \mathbf{J}}{\partial t} = \varepsilon_0 \omega_p^2 \mathbf{E} - \gamma \mathbf{J} + \frac{e}{m}(\rho \mathbf{E} - \mathbf{J} \times \mathbf{B}) + \nabla \cdot \left( \frac{1}{\rho + \varepsilon_0 m \omega_p^2 / e} \mathbf{J}\mathbf{J} \right) \tag{7}$$

$$\frac{\partial \rho}{\partial t} - \nabla \cdot \mathbf{J} = 0, \tag{8}$$

where $\omega_p = \sqrt{e^2 n_0 / \varepsilon_0 m}$ is the plasma frequency. Finally, Eqs. (7) and (8) can be reduced to one equation by substituting $\rho = \varepsilon_0 \nabla \cdot \mathbf{E}$ into them

$$\begin{aligned}\frac{\partial \mathbf{J}}{\partial t} &= \varepsilon_0 \omega_p^2 \mathbf{E} - \gamma \mathbf{J} + \frac{e}{m}\left[\varepsilon_0 (\nabla \cdot \mathbf{E})\mathbf{E} - \mathbf{J} \times \mathbf{B}\right] \\ &+ \nabla \cdot \left( \frac{1}{\varepsilon_0 (\nabla \cdot \mathbf{E}) + \varepsilon_0 m \omega_p^2 / e} \mathbf{J}\mathbf{J} \right)\end{aligned}. \tag{9}$$

In order to avoid solving a large matrix with an implicit difference method, Eq. (9) is solved by two-step splitting scheme [24]. The coupled Maxwell-hydrodynamic system of Eqs. (1), (2) and (9) are numerically solved with the finite-difference time-domain (FDTD) method. A computational grid based on the standard staggered Yee cells is proposed, as shown in Fig. 1. The spatial-temporal dependent **E**, **H,** and **J** are nodally uncollocated in space and staggered in time. The electric field **E** is defined at the time step $l+1/2$ and is located at the cell face center. Both current density **J** and magnetic field **H** are defined at the time level $l$ and are located at the cell face centers and the edges, respectively. This grid arrangement captures the properties of fields and charges as well as the cross-coupling effects between free electrons and electromagnetic fields.

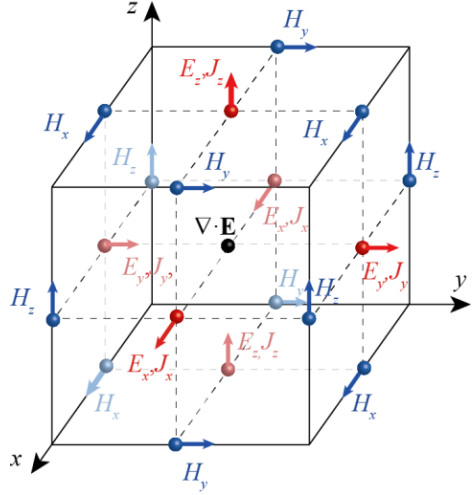

Fig. 1. The Yee grids for the Maxwell-Hydrodynamic system

## 3. Results and Discussion

*3.1 Metasurface THz emitter*

To investigate the mechanism of THz generation from optical metamaterials, we consider an optical metasurface comprising a single layer of SRR array as shown in Fig. 2(a). The thickness of the SRR is 40 nm with a square periodicity of 382 nm. The dimensions of the SRRs are chosen as shown in Fig. 2(b), for which the magnetic dipole can be excited around 200 THz frequency (see the spectra in Fig. 2(c)). The metasurface is placed on a 500-nm-thick glass substrate with the relative permittivity of 2.25. Parameters for gold SRRs are set as $n_0 = 5.92 \times 10^{28}$ m$^{-3}$, $\gamma = 10.68 \times 10^{13}$ rad/s. An *x*-polarized infrared laser pulse propagating in the z-direction is introduced by using the total-field and scattered-field (TF/SF) technique [20]. We describe the pump laser by a Gaussian pulse of the form $E(t) = E_0 \exp\left[-2\ln 2(t-t_0)^2 / \tau^2 \right]\cos(\omega t)$. Here, the driving frequency, temporal width and peak amplitude are chosen as $\omega$ = 1.257×10$^{15}$ rad/s, $\tau$ = 140 fs and $E_0$ = 2×10$^7$ V/m, respectively. PMLs are applied in the *z*-direction and periodic boundaries are employed in both the *x*- and *y*- directions. The temporal evolutions of the transmitted fields were recorded in the total field region. Uniform spatial steps $\Delta x = \Delta y = \Delta z = 2$ nm and time step $\Delta t = 3 \times 10^{-18}$ s are adopted in the simulation.

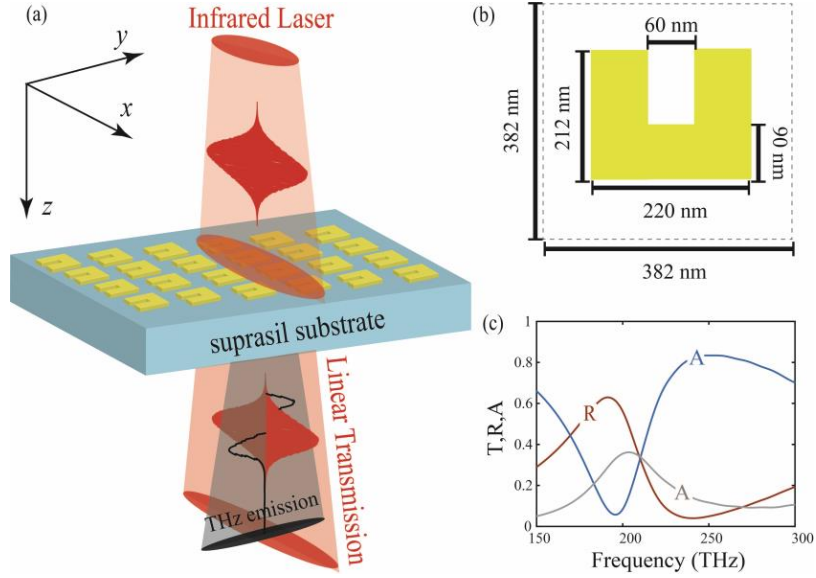

Fig. 2. (a) Schematic of THz generation by illuminating a metasurface with an infrared laser. (b) The feature size of a unit cell. (c) Transmittance T, reflectance R, and absorptance A spectra.

*3.2 Linear and Nonlinear Responses of Metasurface*

Due to the symmetry breaking of SRR, we measure the transmitted electric field $E$ for showing the linear response; and for the second-order nonlinear signal, we record the $E_y$ component. Figures 3(a) and (b) show the incident electric field and nonlinear signal in the time domain. The nonlinear signal is Fourier transformed in the frequency domain exhibiting second-order nonlinear spectra with the two peaks (Fig. 3(c)): the sum-frequency generation has a peak around 400 THz and difference-frequency generation spectrum shows a peak around 0 THz. Here, a femtosecond laser pulse is used to generate THz waves from the SRRs via the optical rectification. Because a femtosecond pulse contains many frequency components, any two frequency components contribute to the difference frequency, and the overall result is the weighted sum of all the contributions. One femtosecond laser pulse is enough to produce the optical-rectification radiation for the THz generation. From the amplitudes of calculated second harmonic and incident pulse, one can roughly estimate the effective nonlinear susceptibility of the metasurface to be

$\chi^{(2)}_{SRR} = \chi^{(1)}_{2\omega} \frac{|E_{2\omega}|}{|E_{inc}|^2} = 1.72 \times 10^{-12}$ m$^2$/V, where $\chi^{(1)}_{2\omega}$ is the linear electric susceptibility of gold at $2\omega$. The calculated $\chi^{(2)}_{SRR}$ value is consistent with the experimental value ~$1.6 \times 10^{-12}$ m$^2$/v in Ref. [25]. In order to extract the THz signal in time domain, the specific frequencies ranging from 0~15 THz are selected and then we inverse Fourier transformed the episode signal back into its time domain counterpart. The extracted THz signal from Fig. 3(b) is plotted in Fig. 3(d) showing a typical time-domain THz trace.

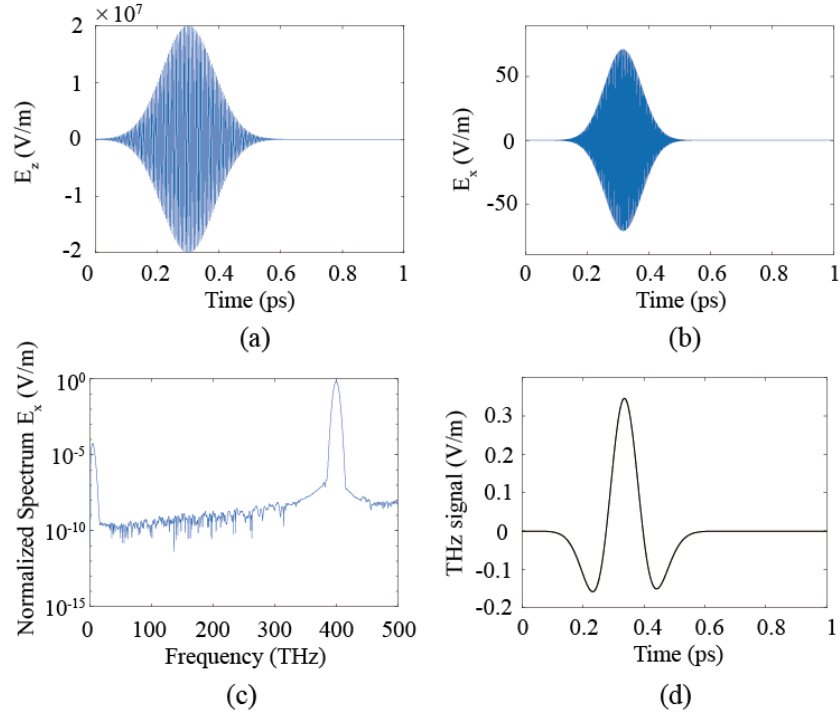

Fig. 3. Electric field of incident pulse $E_{inc}$ (a) and transmitted nonlinear signal Ex (b). (c) Semi-log plot of the Fourier transform of the nonlinear signal in (b). (d) time-domain trace of THz field filtered from the signal in (b).

*3.3 Broadband THz generation*

The single layer metasurface is made of gold and glass, which makes it have very low absorption in the THz frequency region. Thus, the metasurface can be used to generate broadband THz signals. To validate this concept, we first predict the generated THz wave from the metasurface by omitting the thickness of the metasurface since its thickness is just

40 nm, and then comparing the predicted results with the simulation results. The generated THz field $E_{THz}(z,t)$ can be described by the wave equation in a nonlinear medium with the second-order nonlinear susceptibility $\chi^{(2)}$ propagation in the z-axis

$$\frac{\partial^2 E_{THz}(z,t)}{\partial z^2} - \frac{n_{THz}^2}{c^2}\frac{\partial^2 E_{THz}(z,t)}{\partial t^2} = \frac{1}{\varepsilon_0 c^2}\frac{\partial^2 P_{THz}^{(2)}(z,t)}{\partial t^2} = \frac{\chi^{(2)}}{c^2}\frac{\partial^2 |E_0(z,t)|^2}{\partial t^2}, \quad (10)$$

where $n$ is the medium refractive index at the THz frequencies, $c$ is the speed of light, $P_{THz}^{(2)}(z,t)$ the THz polarization which is proportional to the power of the pump laser $|E_0(z,t)|^2$. The pump laser is considered as a temporal Gaussian pulse $E_0(t) = \cos(\omega_0 t + \phi)g_\sigma(t)$. Here $g_\sigma(t) = e^{-\frac{1}{2}\sigma^2 t^2} \Leftrightarrow g_\sigma(\omega) = (\sqrt{2\pi}/\sigma)e^{-\frac{\omega^2}{2\sigma^2}}$ is a Gaussian envelope. $\omega_0$ is the center frequency of the pump pulse and $\phi$ is a phase delay. The Full width at half maximum (FWHM) of the pump pulse is $\Delta f_{FWHM}^{pump} \approx 2.335\sigma$. THz generation from the infinitesimal thin nonlinear metasurface can be expressed as

$$\begin{aligned}P^{(2)}(t) &\sim E^2(t) = \cos(2\omega_0 t + 2\phi)g_\sigma^2(t) + g_\sigma^2(t) \Leftrightarrow \\ P^{(2)}(\omega) &\sim e^{2i\phi}g_{\frac{1}{\sqrt{2}\sigma}}(\omega - 2\omega_0) + e^{-2i\phi}g_{\frac{1}{\sqrt{2}\sigma}}(-\omega - 2\omega_0) + g_{\frac{1}{\sqrt{2}\sigma}}(\omega).\end{aligned} \quad (11)$$

We can see the first two terms are the SHG with a center frequency $2\omega_0$; the last term is the origin of the THz emission; thus, the THz emission only depends on the temporal Gaussian envelope. Consequently, the emitted THz field can be written as

$$\begin{aligned}E^{(THz)}(\omega) &\sim \chi^{(2)}(-i\omega)^2 g_{\frac{1}{\sqrt{2}\sigma}}(\omega) = \chi^{(2)}\omega^2 e^{-\frac{\omega^2}{4\sigma^2}} \Leftrightarrow \\ E^{(THz)}(t) &\sim -\chi^{(2)}\partial_t^2 g_{\sqrt{2}\sigma}(t) = \chi^{(2)}\sigma^2(1 - 2\sigma^2 t^2)e^{-\sigma^2 t^2}\end{aligned} \quad (12)$$

As shown in Fig. 2(c), the metasurface THz emitter resonance is very wide compared to the incident wave bandwidth. Consequently, the effect of the absorption spectrum of metasurface on the THz emission is insignificant. From the Eq. (12), we can see and the bandwidth of THz emission is limited mainly by the duration of the incident pulse. The THz emission has a peak frequency at $f_0^{THz} = 2\sigma$ and a bandwidth $\Delta f_{FWHM}^{THz} = 2.31\sigma$. We can theoretically estimate the peak frequency and bandwidth of the THz emission by the bandwidth of pump pulse, i.e. $\Delta f_{FWHM}^{THz} \approx 0.98\Delta f_{FWHM}^{pump}$ and $f_0^{THz} \approx 0.85\Delta f_{FWHM}^{pump}$. The bandwidth of the generated THz spectrum can be tuned by changing the duration of the incident Gaussian pulse. The control of the central frequency and bandwidth of THz emission spectra by varying the bandwidth of incident pulse is shown in Fig. 4. Figure 4(c) compares the THz emission bandwidth $\Delta f^{THz}$ and central frequency $f_0^{THz}$ from the metasurface versus the incident pulse bandwidth $\Delta f^{inc} \approx 0.44/\tau$. One can see that, THz emission bandwidth and central frequency scale linearly with the pump pulse bandwidth. Comparing with the nonlinear crystals, the single layer metasurface emitter is free from the limitation of the Reststrahlen region [26] and can achieve tunable THz bandwidth by changing the pump pulse duration. Moreover, the fundamental pumping frequency of the metasurface THz emitter is flexible and can easily be scaled to arbitrary operation frequencies by optimizing the dimensions of the SRR unit.

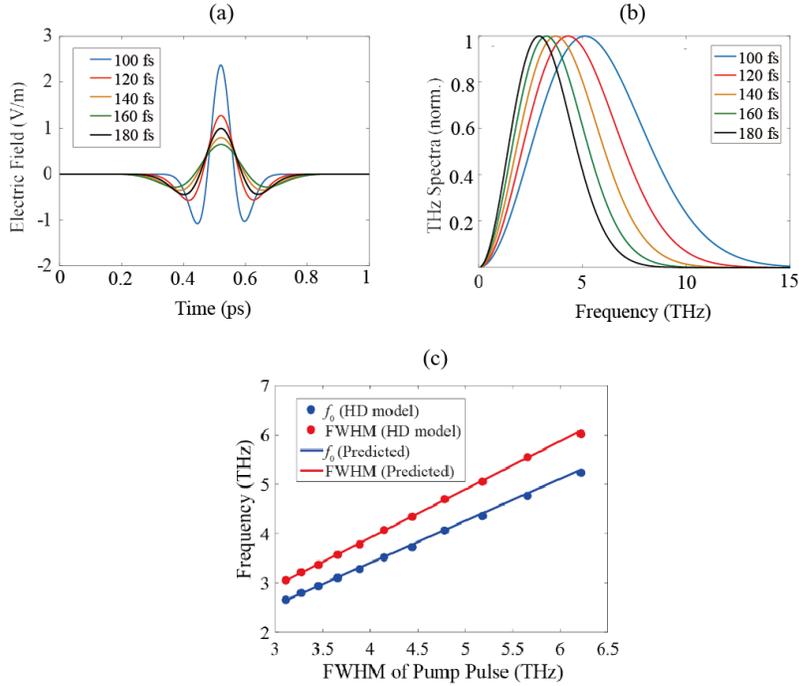

Fig. 4. Comparisons of pump durations and measured spectral bandwidths. (a,b) The THz pulse traces and corresponding normalized spectral amplitudes measured at several pumping durations (100 fs-180 fs). (c) The FWHM of pump pulse versus the generated THz spectral bandwidth (red dots) and peak frequency (blue dots). The straight lines are the predicted analytical results.

*3.4. Polarization Dependence of the metasurface THz emitter*

To further investigate the nature of the metasurface THz emitter, the polarization dependence of the THz amplitude by rotating the pumping polarization is shown in Fig. 5(a). We define the polarization angle to be 0 degrees when the incident pulse is polarized along the *x*-direction (See Fig. 2(a)). Figure 5(a) shows the time-domain THz electric fields for 4 different polarization angles from 0 to 90 degrees. Figure 5(b) shows the polar plot of the peak-to-peak amplitude of THz signals versus the polarization angle. It should be noted that the simulated result can be well fitted with a $\cos^2(\theta)$ function satisfying Lambert's cosine law. Additionally, this polarization control also works for the SHG reported previously [27].

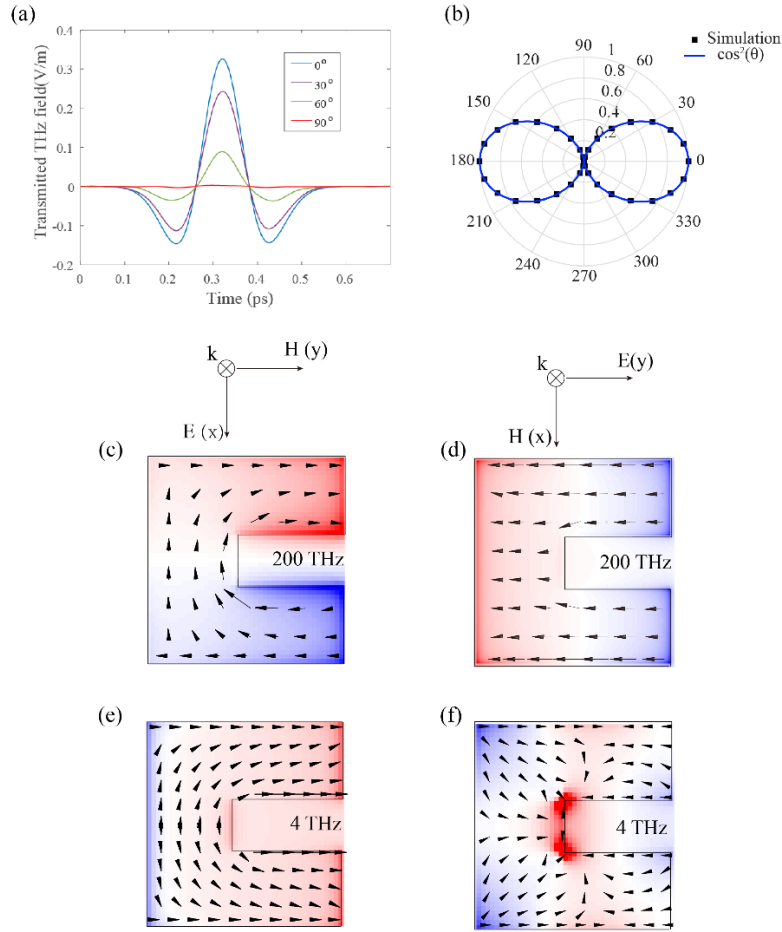

Fig. 5. Polarization dependence of THz generation from the SRRs metasurface. (a) THz signals for different polarization angles. (b) Polar plot of the peak-to-peak amplitude of THz emission as a function of the polarization angle. (c-f) Perpendicular components of the surface electric field (color scale) and current density (arrows) of the unit cell irradiated by the infrared pulse. The current distributions are shown temporally π/2 phase shifted against the charge distribution.

The THz emission originating from optical rectification is induced by the symmetry breaking in the second-order nonlinear current distribution. The convective acceleration term $\mathbf{v}\cdot\Delta\mathbf{v}$ is the key contribution to the second-order nonlinearities [28]. This term behaves like linear current multiplied by linear accumulated density, i.e. $j\rho$. Consequently, the THz nonlinear current is parallel or antiparallel to the linear current. The charge and current distributions at the frequencies of 4 THz and 200 THz were plotted in Figs. 5(c-f). Fig. 5(c) shows the fundamental magnetic dipole resonance for the normal incidence to the SRR with the electric field breaking the symmetry of the SRR. When the polarization of the incident wave is parallel to the gap of SRRs (along the *x*-direction), the nonlinear currents in both arms are parallel along the *y*-direction and thus strong radiate fields can be observed in the far field as depicted in Fig. 5(e). Conversely, due to the symmetry dependent selection rule, the far field from the nonlinear currents flowing along the *x*-direction is canceled. Differently, when the polarization is perpendicular to the gap of SRRs as shown in Fig.5(d), the nonlinear currents show reverse flow directions along the *y*-direction, leading to a vanishing radiation in the far field as shown in Fig. 5(f).

## 4. Conclusion

In conclusion, we demonstrated a new concept of THz emitter based on a single-layer nonlinear metasurface of nanoscale thickness, representing a new platform for revealing artificial magnetism-induced THz generation. The novel THz generation mechanism is numerically investigated by the self-consistent time-domain implementation of the hydrodynamic model. Our numerical results conclude that the nonlinearity arises from the excited magnetic dipole resonance of SRRs and the metasurface shows a high second-order nonlinear susceptibility. The results also suggest that the THz generation scheme is free from phonon limitation and quasi-phase-matching conditions. The generated THz bandwidth can be arbitrarily tuned by the time duration of the excitation pulse. Moreover, the metasurface THz emitter can easily be scaled to arbitrary operation frequencies by optimizing the dimensions of the SRR unit.

**Funding**

This work was supported the US Department of Energy, Office of Basic Energy Science, Division of Materials Science and Engineering (Ames Laboratory is operated for the US Department of Energy by Iowa State University under contract No. DE-AC02-07CH11358). Work at FORTH was supported by the European Research Council under the ERC Advanced Grant No. 320081 (PHOTOMETA). The work was also supported by NSFC (Nos. 61601166, 61701001, 61701003), National Natural Science Fund for Excellent Young Scholars (No. 61722101), Universities Natural Science Foundation of Anhui Province (Nos. KJ2017ZD51 and KJ2017ZD02), and Thousand Talents Program for Distinguished Young Scholars of China.